\documentclass[conference]{IEEEtran}
\IEEEoverridecommandlockouts
\usepackage{cite}
\usepackage{amsmath,amssymb,amsfonts}
\usepackage{algorithmic}
\usepackage{graphicx}
\usepackage{textcomp}
\usepackage{xcolor}
\usepackage{mathtools}
\usepackage{amsmath}
\usepackage{physics}
\usepackage{subcaption}
\usepackage{mathtools}
\usepackage{bm}
\usepackage{gensymb}
\def\BibTeX{{\rm B\kern-.05em{\sc i\kern-.025em b}\kern-.08em
    T\kern-.1667em\lower.7ex\hbox{E}\kern-.125emX}}

\DeclareMathOperator{\sinc}{sinc}
\newcommand{\mbf}[1]{\mathbf{#1}}
\newcommand{\ie}{, i.e., }

\newcommand{\sigmarx}{\mbf{\Sigma}_\text{rx}}

\begin{document}
\title{Uplink Wave-Domain Combiner for Stacked Intelligent Metasurfaces Accounting for Hardware Limitations}
\author{\IEEEauthorblockN{Maryam Rezvani\IEEEauthorrefmark{1}, Raviraj Adve\IEEEauthorrefmark{1}, Akram bin Sediq\IEEEauthorrefmark{2}, Amr El-Keyi\IEEEauthorrefmark{2}}
	\IEEEauthorblockA{\IEEEauthorrefmark{1}\textit{Dept. of Electrical and Computer Engineering},
		\textit{University of Toronto}, 
		Toronto, Canada 
	}
	\IEEEauthorblockA{\IEEEauthorrefmark{2}\textit{Ericsson Canada}, 
		Ottawa, Canada, \\
		emails: \{mrezvani, rsadve\}@ece.utoronto.ca, \{akram.bin.sediq, amr.el-keyi\}@ericsson.com}
}

\maketitle
\begin{abstract}
    Refractive metasurfaces (RMTSs) offer a promising solution to improve energy efficiency of wireless systems. To address the limitations of single-layer RMTS, stacked intelligent metasurfaces (SIMs), which form the desired precoder and combiner in the wave domain, have been proposed. However, previous analyses overlooked hardware non-idealities that significantly affect SIM performance. In this paper, we study the achievable sum-rate of SIM antennas in an uplink scenario, accounting for hardware constraints. We propose a system model that includes noise and hardware effects, formulate a non-convex sum-rate optimization problem, and solve it using gradient ascent and interior point methods. We compare SIMs and digital phased arrays (DPAs) under Rayleigh fading and 3GPP channels with two conditions: equal number of RF chains and equal physical aperture size. Our results show SIMs outperform DPAs under equal number of RF chains but underperform DPAs with equal aperture size.
\end{abstract}

\begin{IEEEkeywords}
	Stacked Intelligent Metasurfaces (SIM), Uplink wireless communications, Wave-domain combiner.
\end{IEEEkeywords}

\section{Introduction}
Future wireless communication networks must meet the rising demand for higher data rates and increased throughput, while also prioritizing sustainability and energy efficiency~\cite{6Gvision}. To achieve higher data rates, extra-large multiple-input multiple-output (MIMO) antennas, also called extra-large digital phased arrays (DPAs), have been introduced as an enhancement over massive MIMO systems. By increasing the number of antennas in a DPA, the signal-to-noise ratio (SNR) improves, reducing the power required by users. However, using larger DPAs lead to higher power consumption at base stations (BSs) due to the numerous power-hungry radio frequency (RF) chains. 

To tackle BS power consumption, the two main approaches are to introduce low complexity signal processing techniques such as~\cite{Hint} or sharing signal processing between digital and analog/wave domains. Hybrid MIMO systems push part of the signal processing to the analog domain using phase shifters~\cite{HeathHYB} while, metasurface (MTS) antennas share signal processing between electromagnetic (EM) wave and digital domains~\cite{Rezvani2023DMA, SIM_ICC, RISDiRenzo}. Our focus here is on MTS antennas.

MTSs, comprising subwavelength unit-cells, enable manipulation of its properties via voltages applied to diodes in each unit-cell. For example, reflective intelligent surfaces utilize MTSs to manipulate wireless channel properties thereby enhancing channel gain. Similarly, deploying refractive MTSs (RMTSs) at BSs provide the ability to form \textit{any desired radiation pattern} at the EM level with higher energy efficiency and lower latency~\cite{RISDiRenzo}.


The effect of the unit-cell on the impinging wave is captured as a weighting factor. Using single-layer  unit-cells, the phase and amplitude of the weighting factor follow Lorentzian modulation, i.e., are coupled~\cite{DMA2017Analysis, pfeiffer2013HMS}. This coupling limits the radiation patterns that can be achieved by the MTS antenna. Achieving phase-only modulation, spanning $[0, 2\pi)$ interval while keeping the amplitude constant, requires a ``stacked unit-cell" design, where each unit-cell comprises at least three reactive impedance sheets~\cite{HMSLookuptable} where the impedance of each layer is controlled by changing the applied voltage to its varactor diode~\cite{phaseonlyTransmissiveMTS}.

Since full control over the radiation pattern requires independent control over amplitude and phase~\cite{balanis2016antenna} of each weighting factor, using only one layer of phase-only RMTS at BS is insufficient to form any desired radiation pattern (e.g., having to null multiuser interference). To tackle this problem, the work in~\cite{SIM_ICC,SIM_JSAC} introduced a \textit{multi-layer} structure called stacked intelligent metasurfaces (SIM) where several phase-only modulated RMTSs are stacked in front of a small DPA and wrapped by absorber walls~\cite{liu2022programmableSIM}. This design is based on diffractive deep neural networks (D2NNs)~\cite{Jarahi2018allSIM}. In SIM antenna, the number of elements (and RF chains) in the DPA is equal to the number of users served by the BS. Also, using a wave-domain (WD) combiner/precoder implemented by the SIM renders the digital combiner redundant, thereby simplifying the required digital signal processing~\cite{SIM_ICC}.
%

Since its introduction, various aspects of SIM-assisted wireless communication have been studied. In~\cite{SIM_ICC}, a joint power allocation and WD precoder algorithm is introduced for downlink communications. Joint precoder and combiner designs for single-user scenarios with SIM antennas at both the transmitter and receiver are explored in~\cite{SIM_JSAC} and~\cite{SIMjoint2}. \cite{SIM_DOA} demonstrates that SIM antennas can perform a 2D discrete Fourier transform over the air.
\cite{hassan2024efficient} shows that SIM-equipped BSs can generate any desired beam pattern. A channel estimation algorithm for SIM networks is proposed in~\cite{SIMCE}. Finally, 
~\cite{SIMISAC} studies SIM’s potential in integrated sensing and communication.

In all these papers, the focus is on downlink wireless communication where the transmitter\ie BS, is equipped with a SIM antenna. This is while the equally important problem of uplink communication where single-antenna users communicate with a SIM-assisted BS is untouched. The uplink problem is a unique one since SIM must effectively distinguish between the weak signal suffered from channel propagation and noise, and while focusing signal on its DPA, also mitigate noise and interference. 

Furthermore, the previous works often assume SIM structures to be lossless\ie a transmission coefficient of one for each RMTS layer. However, signal power losses occur due to copper and dielectric losses, as well as the resonant nature of each unit-cell~\cite{Eleftheriades2018theoryHMS}, making signal detection amidst background noise even more challenging. Also, the current approach to model interlayer EM propagation is based on the Rayleigh-Sommerfeld relation~\cite{SIM_ICC, SIM_DOA, SIM_JSAC, Jarahi2018allSIM} which assumes large dimensions of implemented RMTSs compared to the wavelength and the EM field is not evaluated close to the RMTS~\cite{goodman2005introduction2FourierOptics}. These assumptions do not hold in SIM structures, leading to an unfair comparison with other antenna designs when using the Rayleigh-Sommerfeld relation to model the interlayer propagation. All these issues motivate us to study the SIM-assisted BSs problem in the uplink while taking into account its realistic hardware limitations.

In this paper, we focus on deploying the SIM at the BS, with users equipped with single isotropic antennas. This differs from~\cite{Dai2021TCOM}, where a multilayer RMTS, similar to SIM, is deployed at the user side. While both works address uplink sum-rate optimization, the systems differ fundamentally: we assume the SIM is at the receiver\ie BS, while~\cite{Dai2021TCOM} places it at the transmitter\ie user. Importantly, our model accounts for non-uniform illumination and system losses between RMTS layers, unlike~\cite{Dai2021TCOM}, which considers only path loss. 

\textit{Contributions}: 
In sum, the contributions of this paper are:
\begin{itemize}
	\item We propose an uplink system model for a BS equipped with a SIM antenna which captures the effect of noise and hardware limitations in these systems, and accurately models the EM propagation between RMTS layers.
	\item To design a WD combiner that maximizes achievable sum-rate in the uplink of the SIM antenna, we propose a GA-based algorithm with a variable step size to facilitate the exploration of local optima. Also, we implement an IPM approach and investigate its performance in comparison with the proposed GA-based algorithm.
	\item Our simulation results show that both in Rayleigh fading and realistic 3GPP channels, our proposed GA-based algorithm performance is comparable with IPM and both can improve the performance of SIM antennas with respect to DPA with an equal number of RF chains in single-user and multi-user scenarios. But under equal aperture size, DPA outperforms SIM antenna.
\end{itemize}

\textit{Notation}: Lowercase letteequipped rs denote scalars. We use lowercase and uppercase boldface letters to denote vectors and matrices, respectively. $(\cdot)^H$ denotes conjugate transpose. We represent the vector fields using an arrow, e.g., $\vec{\mbf{A}}$. Furthermore, by $\mbf{A}_{:,n}$ and $\mbf{A}_{m,:}$, we refer to $n$-th column and $m$-th row of $\mbf{A}$, respectively.  Also, $\jmath$ is defined as $\sqrt{-1}$ and $\sinc(x) = \frac{\sin(\pi x)}{\pi x}$. $\mathcal{CN}(\boldsymbol{\mu}, \mathbf{R})$ represents the complex Gaussian distribution with mean $\boldsymbol{\mu}$ and covariance matrix $\mathbf{R}$.


\section{System and SIM Model}
\begin{figure}
	\centering
	\includegraphics[width= 0.6\linewidth]{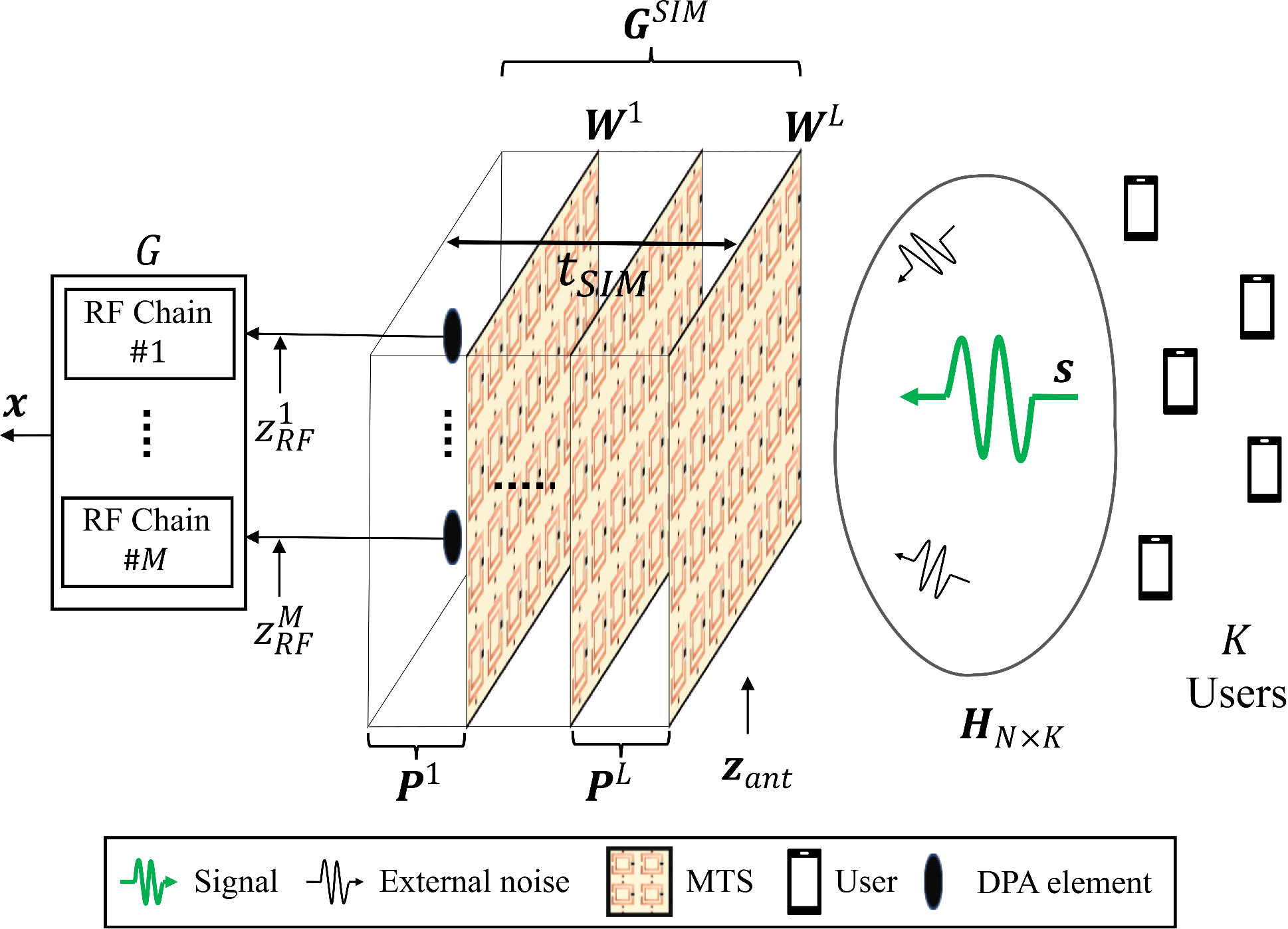}
	\caption{SIM antenna in uplink scenario}
	\label{fig:uplinksys}
\end{figure}
We consider a system where a BS equipped with a $L$-layer SIM, containing $N$ unit-cells in each layer, serves $K$ users, as depicted in Fig.~\ref{fig:uplinksys}. The SIM is placed in front of a $M$-element DPA where each element is connected to an RF chain. The received signal at the BS is described as
\begin{equation}
\small
	\mbf{x} = g(\mbf{G}^\text{SIM}(\sqrt{P_T}\mbf{Hs}+\mbf{z}_\text{ant})+\mbf{z}_\text{RF})
\end{equation}
where $\mbf{x} \in \mathbb{C}^{M \times 1}$ is the received signal at BS, $P_T$ is the transmitted signal power of each user assuming equal power transmission, $\mbf{H} \in \mathbb{C}^{N \times K}$ is the channel matrix, $\mbf{z}_\text{ant} \in \mathbb{C}^{N \times 1}$ is external noise captured by the antenna from the environment; its entries are correlated with same the correlation matrix as the antenna~\cite{Rezvani2023DMA}. Furthermore, $g$ and $\mbf{z}_\text{RF}\in \mathbb{C}^{M\times 1}$ are gain and noise terms introduced by the RF chain where $\mbf{z}_\text{RF}\sim \mathcal{CN}(0,\sigma^2_\text{RF})$, respectively. Finally, $\mbf{G}^\text{SIM} \in \mathbb{C}^{M \times N}$ is the equivalent WD combiner introduced by the SIM structure. In SIM antennas, there is no need for a digital combiner~\cite{SIM_ICC}, hence, we set $M=K$.

Similar to~\cite{SIM_ICC,SIM_JSAC}, we model the effect of the SIM, i.e., $\mbf{G}^\text{SIM}$, as a product of propagation, $\mbf{P}^l$, and RMTS weighting matrices, $\mbf{W}^l$, corresponding to the $l$-th RMTS layer. In the $l$-th layer $\mbf{W}^l$ is a diagonal matrix\ie 
\[ \mbf{W}^l = \text{diag}(e^{\jmath\theta^l_1}, e^{\jmath\theta^l_2}, \dots, e^{\jmath\theta^l_N}) \] 
where $\theta_i^l$ is the phase shift introduced by $i$-th unit cell in $l$-th layer. Likewise, $\mbf{P}^l$ captures the effect of EM propagation between $l$-th and $(l+1)$-th layer.

To model the propagation from DPA to the first RMTS, one common practice is to use Rayleigh-Sommerfeld diffraction theory~\cite{SIM_ICC}. The theory considers EM fields as scalars and is only applicable when the diffracting aperture is large compared to the wavelength and the diffracted field is not observed close to the aperture~\cite[Ch.~3.1]{goodman2005introduction2FourierOptics}. However, in SIM antennas, neither condition is satisfied; importantly, the DPA and RMTSs are in a few wavelengths or even subwavelength distances from each other. For example, denoting $\lambda_0$ as wavelength, in~\cite{SIM_ICC}, the interlayer distance is $5\lambda_0/L$ for $L$ up to $7$. On the other hand, in~\cite{Jarahi2018allSIM}, Rayleigh-Sommerfeld theory is used to model interlayer EM propagation for the D2NN SIM structure since the interlayer distance is at least $5.5\lambda_0$, respectively. We emphasize that accurately modeling the propagation matrix, $\mbf{P}$, is essential to capture variations in EM power due to propagation within the SIM structure and compare SIM antennas with other antenna types fairly.

To model the propagation matrix from the first RMTS layer to DPA, we use the reciprocity theorem and model EM propagation in the reverse direction~\cite[Ch.~7.5]{balanis2012advanced}. In general, the propagation matrix from the DPA to the first RMTS layer depends on the type of antenna elements and their corresponding Green's function~\cite{HMSonCavity}. Here, to provide a valid and illustrative instantiation, we assume that the elements of the DPA are Hertzian dipoles placed in front of a perfect magnetic conductor\footnote{This is a common practice to make the dipole radiation more directive and use the added radiation gain due to the image theory~\cite[Ch.~7.4]{balanis2012advanced}.}. We model variations in the amplitude of impinging EM wave, by capturing relative changes in EM power. For a Hertzian dipole, the Poynting vector, in both near-field and far-field, is described as~\cite{balanis2016antenna}
\begin{equation}
\small
	\begin{split}
		\vec{\mbf{W}}&= \frac{\eta_0 }{8} \abs{\frac{I_0 l_d}{\lambda_0 }}^2 \frac{\sin^2{\theta}}{r^2}\left[1-\jmath \frac{1}{(k_0 r)^3}\right]\hat{r}\\
		&\hspace*{0.5in}+\jmath \eta_0  \frac{k_0 |I_0 l_d|^2\cos{\theta}\sin{\theta}}{16 \pi^2 r^3} \left[1+\frac{1}{(k_0 r)^2}\right]\hat{\theta}\label{eq:dipoleW}
	\end{split}
\end{equation}
where $k_0 = \frac{2 \pi}{\lambda_0}$ is the free space wavenumber, $I_0$ is the excitation current, $l_d$ is the dipole length, $\eta_0 $ is the free space wave impedance, and $r$ and $\theta$ are the distance and azimuth angle between source and evaluation point, respectively. In this paper, we are dealing with uplink communication where BS is in the far field of the users; therefore, only the propagating term of the Poynting vector\ie its real part, is of importance.

Since the relative change in EM power due to propagation is important, we only take into account the term which depends on observation location\ie $\frac{\sin^2{\theta}}{r^2}$. Then, by normalizing this term, we make sure that the total power on a half-sphere at distance $r$ remains equal to the power radiated from the source\footnote{There are no propagation losses from the dipole antenna to the first RMTS layer since the propagation environment is equivalent to free space.}. Further, by mapping the power coefficient to the complex amplitude of $\vec{\mbf{E}}$, we reach the following relation for the propagation of the electric field from the $m$-th element of the DPA to the $n$-th unit-cell on the first RMTS layer 
\begin{equation}
	|\mbf{P}^1_{m,n}| = \left(\frac{3\sqrt{2}}{8\pi}\frac{\sin^2{\theta_{m,n}}}{r^2_{m,n}}\right)^{\frac{1}{2}}.
	\label{eq:PHMSamp}
\end{equation}
where $\theta_{m,n}$ and $r_{m,n}$ are the zenith angle and distance between the $m$-th element of the DPA to the $n$-th unit-cell on first RMTS layer.

To capture the change in phase of the electric field with respect to the source, we calculate the change in phase of the propagating component of the electric field radiated by a Hertzian dipole~\cite[Ch.~4.2]{balanis2016antenna}\ie
\begin{equation}
\small
	\measuredangle \mbf{P}^\text{1}_{m,n} = \frac{\pi}{2}- k_0 r_{m,n}+\measuredangle\left(1+\frac{1}{\jmath k_0 r_{m,n}}-\frac{1}{(k_0 r_{m,n})^2}\right).
	\label{eq:PHMSphase}
\end{equation}
Finally, the $(m,n)$-th element of the propagation matrix $\mbf{P}^1$ is given by $|\mbf{P}_{m,n}^1|e^{\jmath \measuredangle \mbf{P}_{m,n}^1}$. 

The propagation matrix from the $n$-th unit-cell on the $l$-th layer to the $n'$-th unit cell on the $(l+1)$-th layer can be calculated using the same methodology. This is because each unit-cell can be modeled as a dipole~\cite[Ch.~3.3.2]{achouri2021EMMTS} and RMTSs are designed to have zero backward radiation; hence, the sum of the transmission power coefficients on the half sphere in front of each unit-cell should be unity.

By concatenating the propagation matrices and weighting factor of each RMTS, the overall WD combiner of $L$-layer SIM is given by
$\mbf{W}^\text{SIM} = \mbf{P}^1 (\prod_{l=1}^{L-1}\mbf{W}^l\mbf{P}^{l+1})\mbf{W}^L
\label{eq:SIMmodel}$.

\textit{Insertion Loss}: Although RMTSs on each layer are defined as passive and lossless, there are losses due to the propagation of the EM field in the copper (unit-cells) and dielectric. Also, the resonant nature of the unit-cells adds to the losses in each RMTS layer~\cite{Eleftheriades2018theoryHMS}. To account for these losses, we introduce the transmission efficiency coefficient, $T^\text{SIM}$, to the SIM model. Therefore, the WD combiner for $L$-layer SIM is described as:
$\mbf{G}^\text{SIM} = \sqrt{(T^\text{SIM})^L}\mbf{W}^\text{SIM}$.

\section{Problem Formulation and Combiner Design}
The achievable sum-rate, $R$, in the uplink SIM-assisted wireless communication systems can be written as
$R = \sum_{k=1}^K \log_2(1+\gamma_k)$
where $\gamma_k$ is the signal-to-interference-plus-noise ratio for user $k$ defined as
\begin{equation}
    \small
  \frac{P_T\abs{\mbf{G}^\text{SIM}_{k,:}\mbf{h}_{:,k}}^2}{\sum_{\jmath\neq k}^K P_T\abs{\mbf{G}^\text{SIM}_{k,:}\mbf{h}_{:,j}}^2 +\sigma_\text{ant}^2(\mbf{G}^\text{SIM}_{k,:}\mbf{\Sigma_\text{rx}}(\mbf{G}^\text{SIM}_{k,:})^H)+\sigma_\text{RF}^2}.
		\label{eq:gammaSIM}
\end{equation}
We note the two noise terms in the denominator; while the second is the usual noise term in the RF chains, the first term is due to the antenna noise and is colored by the SIM combiner.
To maximize $R$, we are interested in solving the following optimization problem:
\begin{subequations}
	\label{eq:SIMCVX}
        \small
	\begin{align}
		&\max_{\bm{\theta}^1,\dots, \bm{\theta}^L} R\label{OP:obj}\\
		\text{s.t. }& \mbf{W}^\text{SIM} = \mbf{P}^1 \left(\prod_{l=1}^{L-1}\mbf{W}^l\mbf{P}^l\right)\mbf{W}^L\label{OP:cnstr1}\\
		&\mbf{W}^l = \text{diag}(e^{\jmath \bm{\theta}^l}),\label{OP:cnstr2}
	\end{align}
\end{subequations}
where $\bm{\theta}^l = [\theta^l_1, \theta^l_2, \dots, \theta^l_N]^T$ is the vector of phase shifts introduced by the unit-cells.

Solving~\eqref{eq:SIMCVX} is not trivial since $L>1$ and, since the noise is colored, a simple matched-filtering solution is not available even for the single-user case. Hence, we propose a GA-based algorithm comprising the following steps: we begin by randomly initializing $\bm{\theta}^l \ \forall l\in\{1,2,\dots, L\}$ and calculating gradient of~\eqref{OP:obj} with respect to $\theta_l^n$ using:
\begin{subequations}
	\label{eq:SIMGAderivatives}
	\small
	\begin{align}
		&\pdv{R}{{\theta^n_l}} = \sum_{k=1}^K\text{Re} \{\frac{1}{\log(2)} \frac{1}{1+\gamma_k} \pdv{\gamma_k}{{\phi^n_l}}\pdv{{\phi^n_l}}{{\theta^n_l}}\}\\
		&\pdv{\gamma_k}{{\phi^n_l}} = \frac{(\pdv{S_\text{pwr,k}}{{\phi^n_l}}Z_\text{pwr,k} - \pdv{Z_\text{pwr,k}}{{\phi^n_l}}S_\text{pwr,k})}{Z_\text{pwr,k}^2}\\
		\begin{split}
		&\pdv{S_\text{pwr,k}}{{\phi^n_l}} =\\ &\left[\mbf{P}^1_{k,:}\prod_{ll=1}^{l-1}(\mbf{W}^{ll}\mbf{P}^{ll})\right]_{:,n} \left[\prod_{ll = l+1}^{L}(\mbf{P}^{ll}\mbf{W}^{ll})\mbf{H}_{:,k}\right]_{n,:}\mbf{H}^H_{:,k}\mbf{W}^H_{k,:}
		\end{split}\\
		\begin{split}
			&\pdv{Z_\text{pwr,k}}{{\phi^n_l}} = \\ &\sum_{j\neq k}^K \left ([\mbf{P}^1_{k,:}\prod_{ll=1}^{l-1}(\mbf{W}^{ll}\mbf{P}^{ll})]_{:,n} [\prod_{ll = l+1}^{L}(\mbf{P}^{ll}\mbf{W}^{ll})\mbf{H}_{:,j}]_{n,:}\mbf{H}^H_{:,j}\mbf{W}^H_{k,:} \right)\\ &+[\mbf{P}^1_{k,:}\prod_{ll=1}^{l-1}(\mbf{W}^{ll}\mbf{P}^{ll})]_{:,n} [\prod_{ll = l+1}^{L}(\mbf{P}^{ll}\mbf{W}^{ll})\mbf{U}^H]_{n,:}\mbf{U}\mbf{W}^H_{k,:}
		\end{split}
	\end{align}
\end{subequations}
where $\mbf{U}$ is the Cholesky decomposition of the antenna correlation matrix $\mbf{\Sigma}_{rx}$, $\gamma_k=\frac{S_\text{pwr,k}}{Z_\text{pwr,k}}$, and $\pdv{{\phi^n_l}}{{\theta^n_l}} = \jmath e^{\jmath \theta_l^n}$.

Then by normalizing the gradient of each layer, we prevent the vanishing or exploding gradient effect\ie $\pdv{R}{{\theta^n_l}} \leftarrow \frac{1}{\rho_l}\pdv{R}{{\theta^n_l}}$ where $\rho_l$ is the maximum gradient length in the $l$-th layer. Furthermore, the step size $\mu$ is calculated using the \textit{two-way} backtracking line search algorithm~\cite[Ch.~9.2]{boyd2004convex} with $\alpha$ as the initial guess for the step size in each iteration. We look for a proper step length in \textit{both directions} of the gradient to minimize the chance of ending up in a local optimum. Finally, all phase shifts are updated simultaneously using: $\theta_l^n \leftarrow \theta_l^n+\mu\pdv{R}{{\theta^n_l}}$. This process is repeated until some criteria is met.

As both~\eqref{OP:obj} and~\eqref{OP:cnstr2} are non-convex functions, the GA algorithm can converge to a local optimum; at attempt to mitigate the issue we use IPM. The direction of update in each step in IPM is calculated by considering both Hessian and gradient of~\eqref{OP:obj}~\cite{boyd2004convex} and hence, a better update direction is chosen. We use Matlab's \textit{fmincon}~\cite{matlab} to find a sub-optimal WD combiner using IPM with approximate Hessian and gradient. We emphasize that using IPM does not eliminate the possibility of converging to a local optimum.

\section{Numerical Results}
In this section, we document the effectiveness of the proposed GA algorithm and IPM in calculating WD combiner for SIM antenna in single-user and multi-user uplink scenarios where the BS is in the far-field of user(s). Also, as upper and lower performance limits, we report the achievable sum-rate of DPA with an equal physical aperture area as SIM antenna\ie the averaged users' signal power which reaches both antennas are the same, and with equal number of RF chains\ie equal power consumption\footnote{We ignore the power consumption of varactor diodes and driving circuitry of each RMTS unit-cell as being comparatively negligible.}.

We consider an outdoor wireless communication problem, where $K$ users communicate with a BS equipped with an antenna of size $4\lambda_0 \times 4\lambda_0$ at $3$ GHz carrier frequency with $20$ MHz bandwidth. As BS is in the far-field of users, the transmitted power, $P_T$, is defined as the amplitude of the Poynting vector of the users' transmitted signal\ie $P_T = \frac{\abs{\vec{\mbf{E}}_\text{user}}^2}{2\eta_0}$ $[\frac{W}{m^2}]$. The SIM thickness, $t_\text{SIM}$ is $5\lambda_0$\ie the interlayer distance is $5\lambda_0/L$. Furthermore, the RF chains are assumed to be at room temperature\ie $T_\text{BS} = 290 \degree\ [K]$. Finally, $g$ and noise figure of RF chain are set to be $12.5$ [dB] and $18.8$ [dB]\footnote{Calculated by considering a low pass filter~\cite{LPF} and first three elements in the RF chain\ie low noise amplifier~\cite{LNA}, mixer~\cite{mixer}, and in phase/quadrature demodulator~\cite{IQdem}.}, respectively. We repeat the simulation for 10 instances of user placements and 100 channel realizations per instance.  Also, we allow a maximum of 100 and 500 iterations for the IPM and GA algorithms, respectively.

\textit{Received signal power}: The power captured by each antenna/unit-cell is equal to $P_T A_\text{eff}$ where $A_\text{eff}$ is the effective area of the antenna~\cite[Ch.~2.15]{balanis2016antenna}. In the SIM, each unit-cell is approximated by a continuous current sheet, hence, $A_\text{eff}$ is equal to the physical area of each unit-cell $A_p$. Considering unit-cells of size $\frac{\lambda_0}{4} \times \frac{\lambda_0}{4}$ at $3$ GHz, $A_\text{eff,UC} = 0.0025$. For massive MIMO, assuming patch antenna as the elements of DPA, $A_\text{eff}$ is calculated by expressions in~\cite[Ch.~2.16]{balanis2016antenna} that results in $A_\text{eff,PA}$ at $3$ GHz to be $0.0026$.

\textit{Antenna noise power}: The noise power captured by the antenna is defined by 
$P_{z_\text{ant}} = KT_EB\frac{A_\text{eff}}{\lambda_0^2}\Delta \Omega$~\cite[Ch.~13.7.1]{bhattacharyya2006phased} 
where $K$ is the Boltzmann constant, $T_E$ is the environment temperature well approximated by $290 \degree \ K$~\cite{Stutzman}, and $\Delta\Omega$ is the solid angle of noise source at the antenna equal to $2 \pi$.

\textit{Channel model}: In rich scattering environments, we use the Kronecker Rayleigh fading model where $\mbf{H} = \mbf{\Sigma}_{rx}\tilde{\mbf{H}}\mbf{\Sigma}_{tx}$ and $\mbf{\Sigma}_{tx}$ and $\mbf{\Sigma}_{rx}$ are the users' and antennas' correlation matrices, respectively. The entries of $\tilde{\mbf{H}}$ are independent identically distributed (i.i.d.) random variables drawn from a standard normal distribution\ie $\tilde{\mbf{H}} \sim \mathcal{CN}(\mbf{0},\mbf{I}_{NK})$. We assume users are independent of each other hence, $\mbf{\Sigma}_{tx} = \text{diag}(\beta_1, \dots, \beta_K)$ where $\beta_k$ is the path loss of the $k$-th user, here computed using the Winner II urban microcell non-line of sight scenario~\cite{Winner}. To calculate $\mbf{\Sigma}_{rx}$, we assume scatterers are distributed in 3D space and hence, $(\sigmarx)_{(n,n')} = \sinc\left(\frac{2d_{n,n'}}{\lambda_0}\right)$~\cite{paulraj2003introduction} where $d_{n,n'}$ is the distance between the $n$-th and $n'$-th unit-cells. 

In both Rayleigh and 3GPP channels~\cite{quadriga}, since the ground occupies whole view of the antenna, the external noise correlation matrix is equivalent to the antenna correlation matrix in a rich scattering environment\ie $\mbf{z}_\text{ant}\sim \mathcal(0,P_{z_\text{ant}} \sigmarx )$. Also, whenever the antenna's correlation matrix is needed, $\mbf{\Sigma}_{rx}$ is calculated as described above.

Finally, we assume the DPA is formed by an array of patch antennas with an efficiency of $0.9$~\cite{Patchantenna} and the efficiency of each RMTS in SIM, $T^\text{SIM}$, is set to $0.7$~\cite{phaseonlyTransmissiveMTS}. For the sake of comparison, we report a simple matched-filtering combiner, where each $\theta_i^l$ compensates for the phase of the equivalent channel in front of $l$-th layer and $[\mbf{P}^l]_{n,n}$.

\begin{figure}[!t]
	\centering
	\begin{subfigure}{0.45\linewidth}
		\centering
		\includegraphics[height=3.35cm]{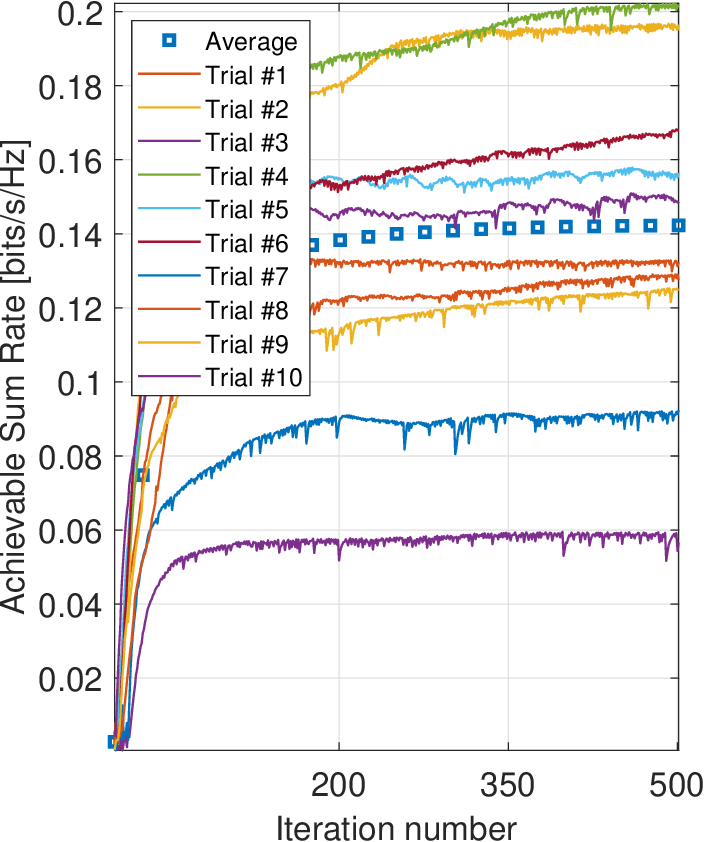}
		\caption{}
		\label{fig:SIMLR}
	\end{subfigure}
	\hfill
	\begin{subfigure}{0.45\linewidth}
		\centering
		\includegraphics[height=3.4cm]{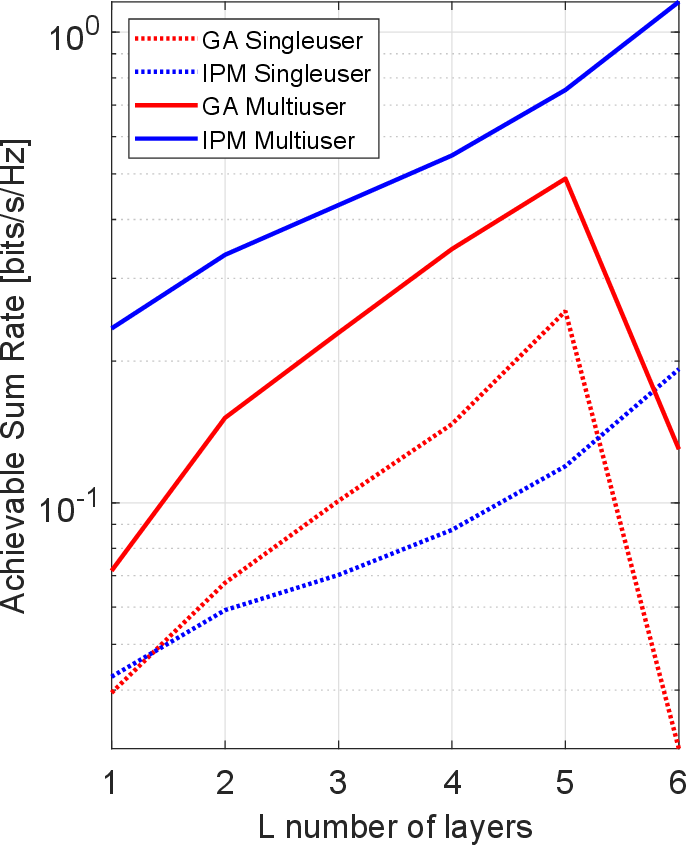}
		\caption{}
		\label{fig:SIMvsL}
	\end{subfigure}
	\caption{SIM (a) Efficiency of the proposed GA algorithm $P_T = 20\ [\text{dBm}/\text{m}^2]$, $L=4$ and $K=1$, (b) SE vs. Number of layers $P_T = 20\ [\text{dBm}/\text{m}^2]$. }
	\label{fig:SIMsim}
\end{figure}

\noindent \textit{Algorithm Efficiency:} We begin by studying the efficiency of the proposed GA algorithm. Fig.~\ref{fig:SIMLR} plots the results averaged over 1000 and for 10 randomly selected  Rayleigh fading channel realizations. As shown, in all cases the major improvement is during the first 200 iterations and all cases converge to at least a local optimum. Also, Fig.~\ref{fig:SIMvsL} plots the achievable sum-rate of SIM using the GA and IPM algorithms for different numbers of layers. As shown, by increasing the number of layers, the achievable sum-rate of both GA and IPM algorithms experience an initial improvement and eventually saturates or experiences a drop. As the number of layers increases, the number of variables increases as well which results in a higher possibility for GA algorithm to converge to a poor local optimum. IPM, using the Hessian of the objective function, can find a better update direction toward the global optimum and escape smaller local optima. Also, when $L$ increases, the interlayer distance decreases and reduced achievable sum-rate, leading to poor interconnectivity as the radiation pattern of each unit-cell is equivalent to a dipole antenna and is not isotropic. Furthermore, more RMTS layers in the SIM structure lead to higher losses in the system due to the insertion loss of each layer. Therefore, there is a sweet spot for $L$ that leads to the best SIM performance.

\begin{figure}[!t]
	\begin{subfigure}{\linewidth}
			\centering
			\includegraphics[width=0.6\linewidth]{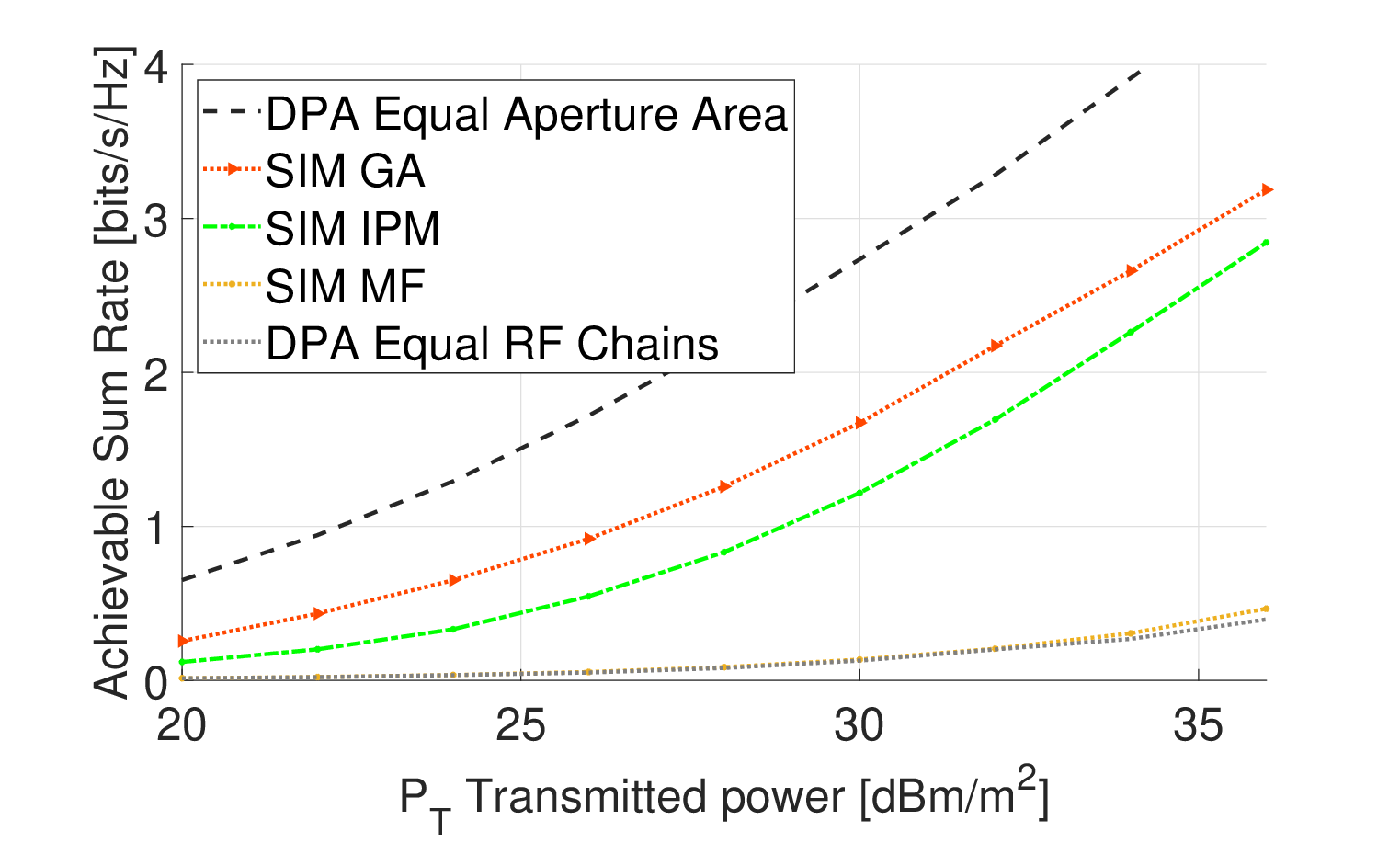}
			\caption{}
			\label{fig:SUR}
	\end{subfigure}
	\hfill
	\begin{subfigure}{\linewidth}
			\centering
			\includegraphics[width=0.6\linewidth]{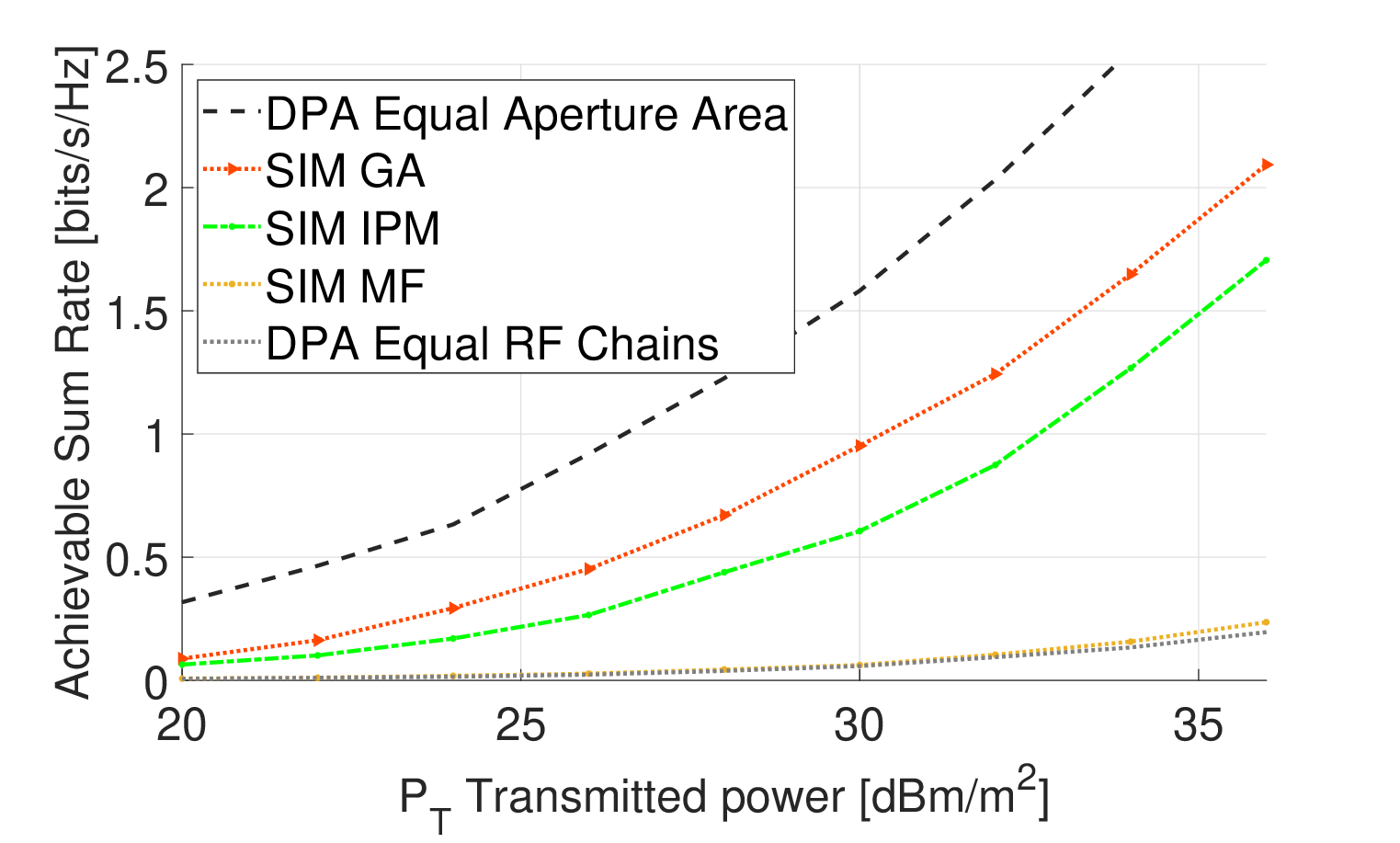} 
			\caption{}
			\label{fig:SUQ}
	\end{subfigure}
	\caption{Achievable sum-rate in (a) Rayleigh fading (b) realistic 3GPP channels. $K=M=1,L=5$. }
	\label{fig:SESU}
\end{figure}

\noindent \textit{Single-User Performance} For a single-user scenario, depicted in Fig.~\ref{fig:SESU} is the achievable sum-rate of a 5-layer SIM, using both IPM and GA algorithms, and DPA with a physical aperture size equal to $4\lambda_0 \times 4 \lambda_0$. Also, for comparison, the achievable sum-rate of a DPA with 
$M=K=1$ RF chains is reported, as is the performance of the SIM with a simple MF combiner. We set $\alpha$ to $1.8$ in the GA algorithm. As shown, using IPM and GA algorithm, we gain a significant improvement in performance compared to using a simple MF combiner. Interestingly, using GA provides a higher sum-rate than using IPM in both Rayleigh fading and realistic 3GPP channels. In this case it appears that the flexibility in choosing the learning rate in each iteration helped GA algorithm in single-user cases to skip local optimums and find a closer-to-optimum combiner than IPM. Also, the achievable sum-rate of SIM antenna is lower than that of the DPA using maximum ratio combiner due to higher losses in the system (in comparison with DPA under equal aperture size constraint). However, when the number of RF chains is made equal, putting a SIM in front of a single antenna \textit{improves the uplink sum-rate of the user}. Here, the SIM acts as a lens that focuses the received signal power by the user on its own receiver antenna.

\begin{figure}[!t]
	\begin{subfigure}{\linewidth}
			\centering
			\includegraphics[width=0.60\linewidth]{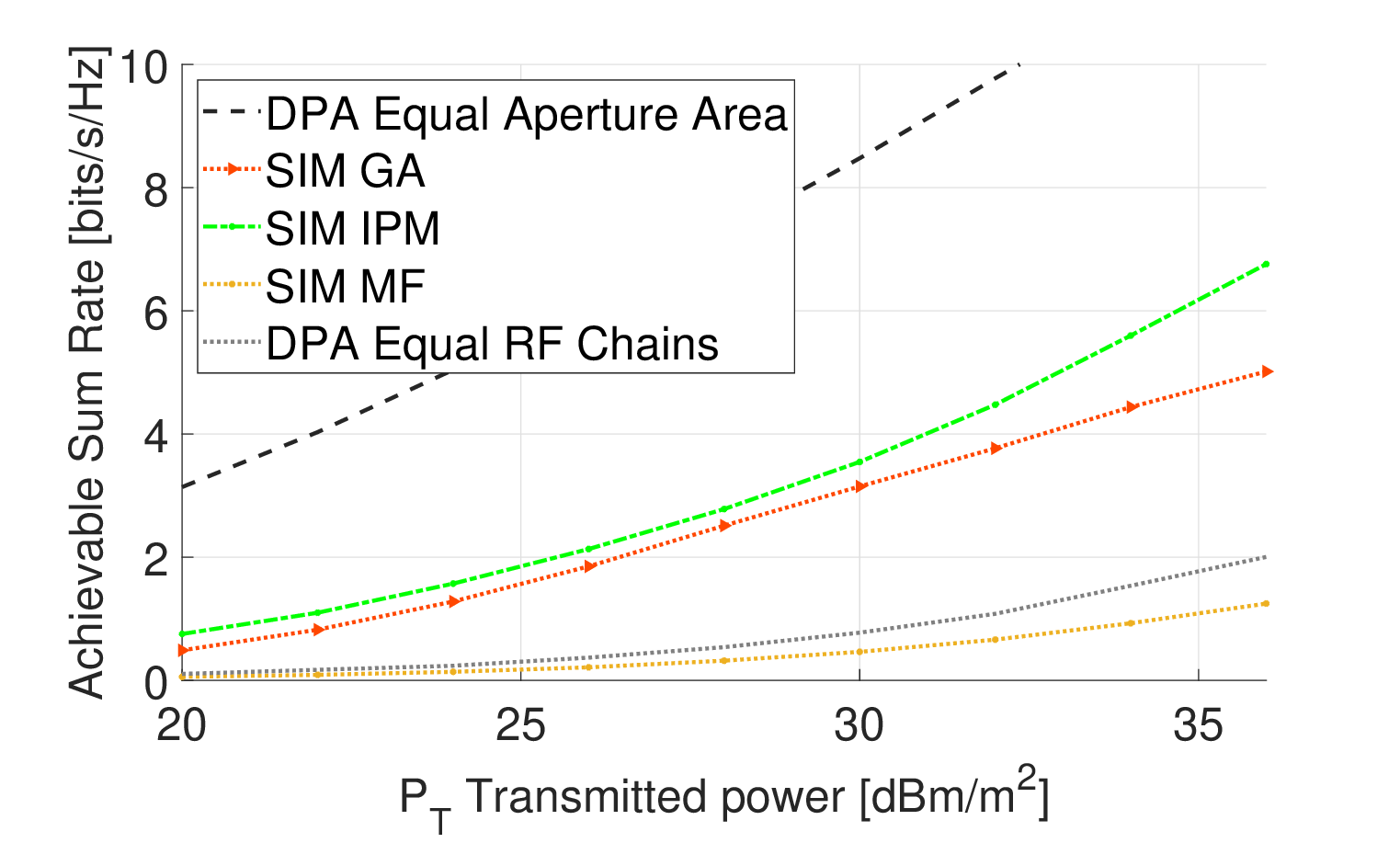}
			\caption{}
			\label{fig:MUR}
	\end{subfigure}
	\hfill
	\begin{subfigure}{\linewidth}
			\centering
			\includegraphics[width=0.60\linewidth]{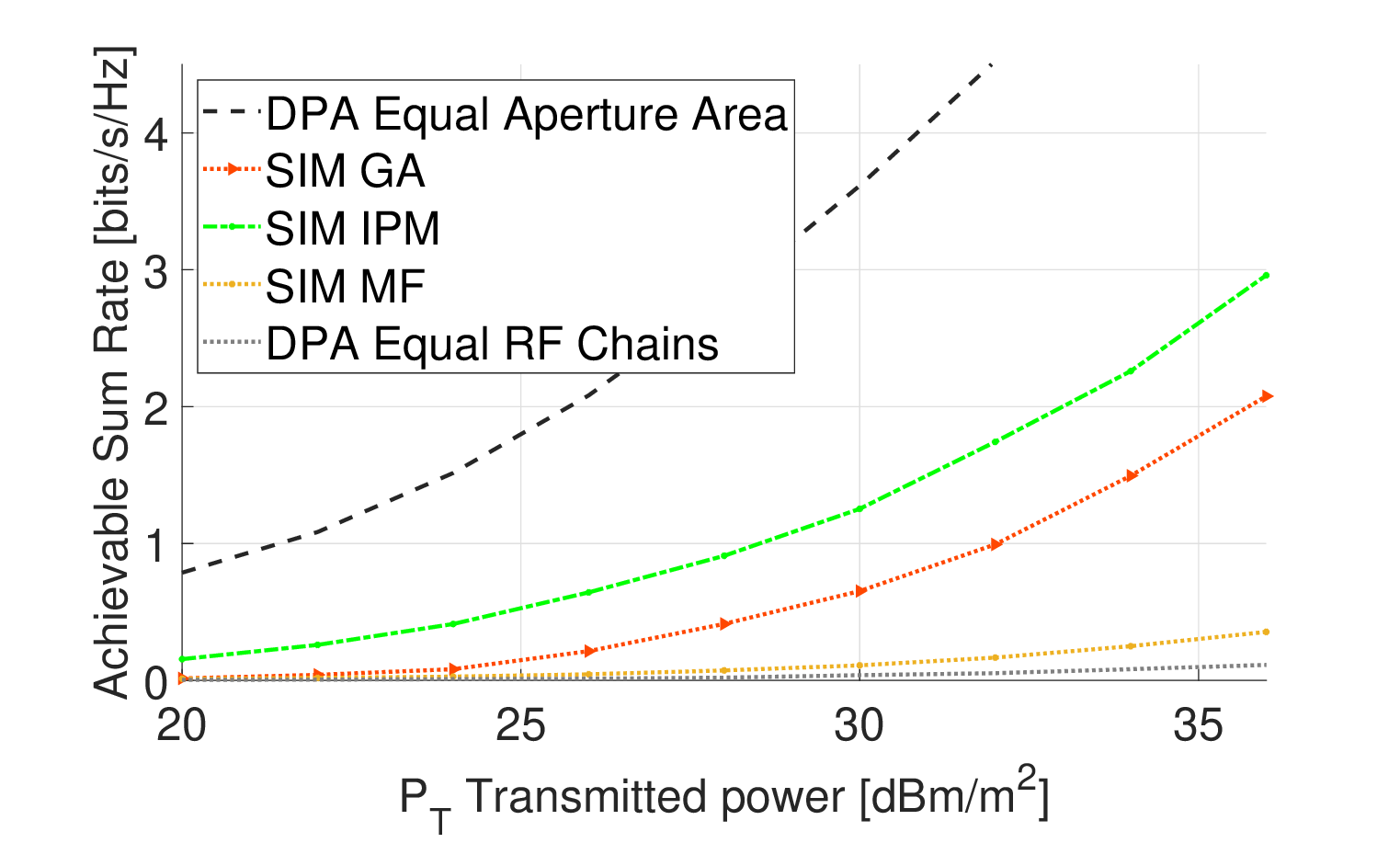} 
			\caption{}
			\label{fig:MUQ}
	\end{subfigure}
	\caption{Achievable sum-rate in (a) Rayleigh fading (b) realistic 3GPP channels. $K=M=2, L=5$. }
	\label{fig:MUSE}
\end{figure}

\noindent \textit{Multiuser Scenario:} Finally, we consider a multi-user scenario with $K=2$, depicted in Fig.~\ref{fig:MUSE}. In this case, DPA uses zero-forcing to combine the received signal, $\alpha$ is set to $2.2$ in the GA algorithm, and we choose a random user to form the MF combiner. As show, in both Rayleigh fading and realistic 3GPP channels, the WD combiner calculated by IPM achieves a higher sum-rate in comparison with GA and MF combiners. The more complex optimization problem due to inter-user interference, increases the possibility of GA algorithm ending up early in a local optimum. It is worth noting that due to the non-convexity of objective function and constraints of~\eqref{eq:SIMCVX}, performance of GA algorithm is highly dependent on the initial guess of the backtracking algorithm, $\alpha$. A more sophisticated algorithm for choosing step length may result in a better performance of GA algorithm.


\section{Conclusion}

In this paper, we explore the use of SIM antennas at the BS for uplink communications. We propose a system model that accounts for noise sources and hardware limitations, and we model interlayer propagation matrices using EM theory to capture signal power variations. We then propose a GA-based algorithm and IPM to design a WD combiner that maximizes the uplink sum-rate. Simulations show that SIM antennas outperform DPAs in Rayleigh fading and 3GPP channels \textit{with equal RF chains}, while DPAs achieve higher sum-rates when both antennas have equal aperture areas.

\section*{Acknowledgments}
This work was supported by the Ericsson Canada. 
Maryam Rezvani gratefully acknowledges insightful discussions with Parham Abbasloo.


\bibliographystyle{IEEEtran}
\bibliography{references}

\end{document}